\documentclass[CEJP,DVI]{cej} % use DVI command to enable LaTeX driver
\usepackage{layout}
\usepackage{amsmath}
\usepackage{textcomp}
\usepackage{hyperref}

\usepackage{epsfig}
\usepackage{graphicx}
\usepackage{amssymb}

\title{Nonextensive statistical effects in the quark-gluon plasma formation at relativistic heavy-ion collisions energies}

\articletype{Research Article}

\author{G. Gervino\inst{1}$^,$\inst{3},
        A. Lavagno\inst{2}$^,$\inst{3},
        D. Pigato\inst{2}$^,$\inst{3}}

\institute{
     \inst{1} Dipartimento di Fisica, Universit\`a di Torino, I-10126 Torino, Italy
     \inst{2} Dipartimento di Fisica, Politecnico di Torino, I-10129 Torino, Italy
     \inst{3} Istituto Nazionale di Fisica Nucleare (INFN), Sezione di Torino, I-10126 Torino, Italy
          }

\abstract{We investigate the relativistic equation of state of hadronic matter and quark-gluon plasma at finite temperature and baryon density in the framework of the non-extensive statistical mechanics, characterized by power-law quantum distributions.  We impose the Gibbs conditions on the global conservation of baryon number, electric charge and strangeness number. For the hadronic phase, we study an extended relativistic mean-field theoretical model with the inclusion of strange particles (hyperons and mesons).  For the quark sector, we employ an extended MIT-Bag model.
In this context we focus on the relevance of non-extensive effects in the presence of strange matter.}


\begin{document}
\maketitle

%% ###################################################################

\section{Introduction}

The behavior of hot, dense nuclear matter is usually described in terms of the so-called relativistic mean field models, wherein thermodynamical quantities are usually obtained through the common Boltzmann-Gibbs statistics. But, in recent years, there have been growing theoretical and experimental indications that, under extreme conditions and during the possible phase transition to quark-gluon matter reachable in high energy heavy ion collisions, strong dynamical correlations, long-range color interactions and microscopic memory effects can take place \cite{tsallis1,tsallis2,tsallis3,wilk,wilk2,kodama1,bediaga,beck1,
biroprl051,biroprl052,biro3,alberico,albe2,plb,physicaA,epj,quon,cley1}.
In this context the non-extensive statistical mechanics proposed by Tsallis \cite{tsallis1,tsallis2,tsallis3} can be used to describe and investigate such physical phenomena.
Nonextensive statistical effects should strongly affect the finite temperature and nuclear density Equation of State (EOS) \cite{tew,njl,pereira1,1lavoro,epj2011}.  In fact, by varying temperature and density, the EOS reflects (in terms of the macroscopic thermodynamical variables) the microscopic interactions between the different nuclear matter phases. This means that the physical proprieties of the system can be sensibly changed with respect to the standard Boltzmann-Gibbs statistics.

Unfortunately, extracting information about the EOS at different densities and temperatures by means of heavy ion collisions is very difficult and can be realized only indirectly by comparing the experimental data with different theoretical models such as, for example, fluid-dynamical models \cite{iva}. It is also relevant that a non-extensive hydrodynamic model for multiparticle production processes has been proposed \cite{wilk08}.

In this paper, we extend previous investigation (Ref. \cite{1lavoro}) to the regime of heavy-ion collisions, by by considering the physical conditions in which strange (hadronic and quark) particles can take place, and we analyze the mixed phase region following the Gibbs condition for the phase equilibrium by requiring the global conservation of the three charges: baryon number, electric charge and zero net strangeness.

We show that, in the presence of non-extensive statistical effects, there is a remarkable variation in many physical features such as a strong enhancement of the strange hadronic particles production both in the hadronic and in the mixed phase, together with a significant variation in the particle concentrations and a softening of the pressure in the mixed phase.

% --------------------------
\section{Nonextensive hadronic equation of state}
Nonextensive statistical mechanics, introduced by Tsallis \cite{tsallis1,tsallis2,tsallis3}, is a generalization of the common Boltzmann-Gibbs statistics, based upon the introduction of the
following entropy
\begin{eqnarray}
S_q[f]=\frac{1}{q-1}\, \left(1-\int[f({\bf x})]^q
\,d\Omega\right)\; ,\ \ \ \left(\int f({\bf x})\,d\Omega=1\right)
\, , \label{eq:GMTsallis}
\end{eqnarray}
where $f({\bf x})$ is a normalized probability distribution, with
${\bf x}$ and $d\Omega$ denoting, respectively, a generic point and
the volume element in the corresponding phase space.
The generalized entropy has the usual properties of positivity,
equiprobability, concavity and irreversibility, and in the limit of
$q\rightarrow 1$, the entropic form
(\ref{eq:GMTsallis}) becomes additive and reduces to the standard
Boltzmann-Gibbs entropy
\begin{equation}
S_1=-\int f({\bf x})\, \ln f({\bf x})\, d\Omega\, .
\end{equation}
A second crucial assumption on non-extensive statistics is the
introduction of the $q$-mean value (or escort mean value) of a
physical observable $A({\bf x})$
\begin{equation}
 \displaystyle\langle A\rangle_q=\frac{\int A({\bf x})\,[f({\bf x})]^q
d\Omega} {\int [f({\bf x})]^q d\Omega} \, . \label{escort}
\end{equation}

The non-extensive statistical effects vanish approaching zero temperature, and such a formalism can be considered as an appropriate basis to deal with physical phenomena in which strong dynamical correlations, long-range interactions and microscopic memory effects take place \cite{tsallis1,tsallis2,tsallis3,wilk}.

The probability distribution can be obtained maximizing the measure $S_q$ under appropriate constraints related to the previous definition of the $q$-mean value. In this context, it is
important to observe that the Tsallis classical distribution can be seen as a superposition of Boltzmann distributions with different temperatures whose mean values correspond to
the temperature in the Tsallis distribution. The non-extensive $q$ parameter is related to the temperature fluctuation and describes the spread around the average value of the Boltzmann temperature \cite{wilk}.
Moreover, let us remember that, in the diffusional approximation, a value $q\neq 1$ implies anomalous diffusion among the constituent particles with superdiffusion if $q>1$, and subdiffusion if $q<1$ \cite{tsamem}.

From the above, we can obtain the associate quantum mean occupation number of particles species $i$ in a grand canonical ensemble. For a dilute gas of particles and for small deviations from the standard statistics ($q\approx 1$), the occupation number can be written as \cite{tew,njl}
\begin{equation}
n_i=\frac{1} { {\tilde{e}}_q(\beta(E_i-\mu_i)) \pm 1} \, ,
\label{eq:distribuzioneq>1a}
\end{equation}
where $\beta=1/T$ and the sign option $(\pm 1)$ is for fermions and
bosons respectively. Furthermore, in Eq.(\ref{eq:distribuzioneq>1a}), following Ref. \cite{njl}, for $q>1$, we have
\begin{equation}
{\tilde{e}}_q(x)=\begin{cases}[1+(q-1)x]^{1/(q-1)} &\text{if} \; x>0 ; \\
[1+(1-q)x]^{1/(1-q)} &\text{if} \; x\le 0 ,
\end{cases} \label{eq:distribuzioneq2}
\end{equation}
whereas, for $q<1$,
\begin{equation}
{\tilde{e}}_q(x)=\begin{cases} [1+(q-1)x]^{1/(q-1)} &\text{if} \; x\le 0 ; \\
[1+(1-q)x]^{1/(1-q)} &\text{if} \; x>0 . \end{cases}\label{eq:distribuzioneq3}
\end{equation}
Naturally, as $q\rightarrow1$, the above quantum distribution reduces to the standard Fermi-Dirac and Bose-Einstein distributions.

Let us observe that when the entropic $q$ parameter is smaller than one, the high-energy tail of the above particle distribution is depleted; whereas when $q$ is greater than one, it is enhanced. Hence the non-extensive statistics entails a sensible difference of the power-law particle distribution shape in the high energy region compared with the standard statistics. From a phenomenological point of view, the non-extensive index $q$ is considered here as a free parameter, even if it is actually not because it should depend, in principle, on the physical conditions generated in the reaction and on the fluctuation of the temperature, and also be related to microscopic quantities (such as the mean inter-particle interaction length, the screening length and the collision frequency into the parton plasma).

In the present investigation we are going to study small deviations from
the standard statistics and values of $q>1$ only, because these
values were obtained in several phenomenological studies of high
energy heavy ion collisions (see, for example, Ref.s \cite{alberico,physicaA,cley1,wilk3}).

In this framework, we analyze hadron interaction through the relativistic
mean field (RMF) model \cite{wal,buguta,glen,ditoro}, in order to investigate the equation of state of nuclear matter at finite baryon density and at temperatures reachable in high energy heavy ion collisions.

The Lagrangian density for the full octet of baryons ($p$, $n$, $\Lambda$, $\Sigma^+$, $\Sigma^0$,
$\Sigma^-$, $\Xi^0$, $\Xi^-$) can be written as
\begin{eqnarray}\label{lagrangian}
{\cal L}_{\rm octet} \!\!\!&=&\!\!\!
\sum_i\bar{\psi}_i\,[i\,\gamma_{\mu}\,\partial^{\mu}-(M_i- g_{\sigma
i}\,\sigma) -g_{\omega i}\,\gamma_\mu\,\omega^{\mu} -g_{\rho i}\,\gamma_{\mu}\,\vec{t}
\cdot \vec{\rho}^{\;\mu}]\,\psi_i
+\,\frac{1}{2}(\partial_{\mu}\sigma\partial^{\mu}\sigma-m_{\sigma}^2\sigma^2)
-U(\sigma)\nonumber\\
&&+\frac{1}{2}\,m^2_{\omega}\,\omega_{\mu}\omega^{\mu}
+\frac{1}{4}\,c\,(g_{\omega N}^2\,\omega_\mu\omega^\mu)^2
+\,\frac{1}{2}\,m^2_{\rho}\,\vec{\rho}_{\mu}\cdot\vec{\rho}^{\;\mu}
-\frac{1}{4}F_{\mu\nu}F^{\mu\nu}
-\frac{1}{4}\vec{G}_{\mu\nu}\vec{G}^{\mu\nu}\,,
\end{eqnarray}
where the sum runs over all baryon octets, $M_i$ is the vacuum
baryon mass of index $i$, $\vec{t}$ denotes the isospin
operator which acts on the baryon and $U(\sigma)$ is the nonlinear
self-interaction potential of $\sigma$ meson
\begin{eqnarray}
U(\sigma)=\frac{1}{3}a\,(g_{\sigma
N}\,\sigma)^{3}+\frac{1}{4}\,b\,(g_{\sigma N}\,\sigma^{4}) \,,
\end{eqnarray}
introduced by Boguta and Bodmer \cite{buguta} in order to achieve a reasonable compressibility at the saturated density of nuclear matter.

The meson fields are coupled with the baryon octet through opportune
model-dependent coupling constants.  In this scheme, the effective baryon mass is given by $M^*_i= M_i- g_{\sigma i}\sigma$, and the scalar
and vector baryon density $\rho^S_i$ and $\rho^B_i$ are given, respectively, by
\begin{eqnarray}
&&\rho^S_i= \gamma_i \int\frac{{\rm
d}^3k}{(2\pi)^3}\,\frac{M_i^*}{E_i^*}\,
[n_i^q(k)+\overline{n}_i^{\,q}(k)] \, ,  \label{eq:rhos} \\
&&\rho^{B}_i= \gamma_i \int\frac{{\rm
d}^3k}{(2\pi)^3}[n_i(k)-\overline{n}_i(k)]\, ,\label{eq:rhob}
\end{eqnarray}
where $\gamma_i=2J_i+1$ is the degeneracy spin factor
and $n_i(k)$ and $\overline{n}_i(k)$ are the $q$-deformed particle and antiparticle distributions function given in Eq.s(\ref{eq:distribuzioneq>1a})-(\ref{eq:distribuzioneq3}); for example for $q>1$ and $\beta(E_i^*-\vert\mu_i^*\vert)>0$, we have
\begin{eqnarray}
n_i(k)=\frac{1} { [1+(q-1)\,\beta(E_i^*(k)-\mu_i^*)
]^{1/(q-1)} + 1} \label{eq:distribuz} \, , \\
\overline{n}_i(k)=\frac{1} {[1+(q-1)\,\beta(E_i^*(k)+\mu_i^*)
]^{1/(q-1)} + 1} \, . \label{eq:distribuz2}
\end{eqnarray}
The baryon effective energy is defined as
$E_i^*(k)=\sqrt{k^2+M_i{^*}^2}$ and the effective chemical
potentials $\mu_i^*$  are given in terms of the meson fields as follows
\begin{eqnarray}
\mu_i^*={\mu_i}-g_{\omega i}\omega - g_{\rho i}\tau_{3i}\rho \, ,
\label{mueff}
\end{eqnarray}
where $\mu_i$ are the thermodynamical chemical potentials
$\mu_i=\partial\epsilon/\partial\rho_i$. At zero temperature they
reduce to the Fermi energies and the non-extensive statistical effects
disappear. The meson fields are obtained as a solution of the field
equations in mean field approximation and the related meson-nucleon
couplings constant ($g_{\sigma N}$, $g_{\omega N}$ and $g_{\rho N}$)
will be fixed to the parameters set marked as GM3 of Ref.\cite{glen}.

Because we are going to describe a finite temperature and density nuclear matter with respect to strong interaction, we have to require the conservation of three `charges': baryon number ($B$), electric charge ($C$) and strangeness number ($S$).  For this reason the system is described by three independent chemical potentials: $\mu_B$, $\mu_C$ and $\mu_S$, respectively, the baryon, the electric charge and the strangeness chemical potentials. Therefore, the chemical potential of particle of index $i$ can be written as
\begin{equation}
\mu_i=b_i\, \mu_B+c_i\,\mu_C+s_i\,\mu_S \, , \label{mu}
\end{equation}
where $b_i$, $c_i$ and $s_i$ are, respectively, the baryon number, the electric charge and the strangeness quantum numbers of the $i$-th hadronic species (baryons and mesons).

The thermodynamical quantities can be obtained from the baryon grand potential $\Omega_B$ in the standard way. More explicitly, the baryon pressure $P_B=-\Omega_B/V$ and the energy density can be written as
\begin{eqnarray}
P_B&=&\frac{1}{3}\sum_i \,\gamma_i\,\int \frac{{\rm d}^3k}{(2\pi)^3}
\;\frac{k^2}{E_{i}^\star(k)}\; [n^q_i(k)+\overline{n}^q_i(k)]
-\frac{1}{2}\,m_\sigma^2\,\sigma^2 - U(\sigma)+
\frac{1}{2}\,m_\omega^2\,\omega^2\nonumber\\
&+&\frac{1}{4}\,c\,(g_{\omega N}\,\omega)^4
+\frac{1}{2}\,m_{\rho}^2\,\rho^2\, ,\\
\epsilon_B&=&\sum_i \,\gamma_i\,\int \frac{{\rm
d}^3k}{(2\pi)^3}\;E_{i}^\star(k)\; [n^q_i(k)+\overline{n}^q_i(k)]
+\frac{1}{2}\,m_\sigma^2\,\sigma^2
+U(\sigma)+\frac{1}{2}\,m_\omega^2\,\omega^2\nonumber \\
&+&\frac{3}{4}\,c\,(g_{\omega N}\,\omega)^4 +\frac{1}{2}\,m_{\rho}^2
\,\rho^2 \, .
\end{eqnarray}
Moreover, as temperature increases and baryon density decreases, the mesonic contribution to the total thermodynamical potential becomes increasingly important.  For simplicity, we treat the mesons as an ideal Bose gas so we evaluate the pressure $P_M$, energy density $\epsilon_M$, particle density $\rho^M$ of mesons as
\begin{eqnarray}
&&P_M= \frac{1}{3}\sum_j \,\gamma_j\,\int \frac{{\rm
d}^3k}{(2\pi)^3} \;\frac{k^2}{E_{j}(k)}\; [n^q_j(k)+\bar{n}^q_j(k)]\, ,
\label{pmeson}\\
&&\epsilon_M=\sum_j \,\gamma_j\,\int \frac{{\rm
d}^3k}{(2\pi)^3}\;E_{j}(k)\; [n^q_j(k)+\bar{n}^q_j(k)] \, ,\label{emeson}\\
&&\rho_j^M=\gamma_j \int\frac{{\rm
d}^3k}{(2\pi)^3}\;[n_j(k)-\bar{n}_j(k)] \, , \label{rhomeson}
\end{eqnarray}
where $\gamma_j=2J_j+1$ is the degeneracy spin factor of the
$j$-th meson and the functions $n_j(k)$ $\bar{n}_j(k)$ are the $q$-deformed
boson particle (antiparticle) distributions of the $j$-th
meson. The total pressure and energy density is given as usual by $P_{tot}=P_B+P_M$ and $\epsilon_{tot}=\epsilon_B+\epsilon_M$.

\section{Quark-gluon equation of state}

In this work we use a simple effective MIT bag model to describe the quark phase. All the non-perturbative effects are simulated by the bag constant $B$ which represents the pressure of the vacuum. It is well known that, using the simplest version of the MIT bag model, at moderate temperatures the deconfinement transition takes place at very large densities if the bag pressure $B$ is fixed to reproduce the critical temperature computed in lattice QCD. On the other hand, there are strong theoretical indications that at moderate and large densities (and not too large temperatures) diquark condensates can form, whose effect can be approximately accounted for by reducing the effective bag constant \cite{pagliara}. As proposed in Ref. \cite{bonanno},  a phenomenological approach can therefore be based on an effective "bag constant" $B_{\rm eff}$ depending on the baryon chemical potential. It can be written as
\begin{eqnarray}
B_{\rm eff}= (B_0 - B_\infty)/(1 + \exp[(\mu_B - \mu_0)/a]) +
B_\infty \, ,
\end{eqnarray}
where we have set $B_0^{1/4}=250$ MeV (bag constant at vanishing
$\mu_B$), $B_\infty^{1/4}=160$ MeV (bag constant at very large
$\mu_B$), $\mu_0=600$ MeV and $a=320$ MeV.
The above parameter values have been fixed by requiring that, at low $\mu_B$, the critical temperature is $\approx 170$ MeV for $q=1$, while the other constraint is the requirement that the mixed phase starts forming at a density slightly exceeding $3\rho_0$ for a temperature of the order of $T\approx 90\div 100$ MeV (as also suggested, e.g., by \cite{toneev}).

Following this line, the pressure, energy density and baryon number density for a relativistic Fermi gas of quarks can be written, respectively, as
\begin{eqnarray}
&P& =\frac{\gamma_f}{3} \sum_{f} \int^\infty_0 \frac{{\rm
d}^3k}{(2\pi)^3} \,\frac{k^2}{e_f}\, [n^q_f(k)+\overline{n}^q_f(k)]
-B_{\rm eff}\,, \label{bag-pressure}\\
&\epsilon& =\gamma_f \sum_{f}  \int^\infty_0 \frac{{\rm
d}^3k}{(2\pi)^3} \,e_f\, [n^q_f(k)+\overline{n}^q_f(k)]
\label{bag-energy}
+B_{\rm eff}\,, \\
&\rho& =\frac{\gamma_f}{3} \sum_{f} \int^\infty_0 \frac{{\rm
d}^3k}{(2\pi)^3}  \,[n_f(k)-\overline{n}_f(k)]\, ,
\label{bag-density}
\end{eqnarray}
where the quark degeneracy for each flavor ($f=u,d,s$) is
$\gamma_f=6$, $n_f(k)$ and $\overline{n}_f(k)$ are the $q$-deformed particle
antiparticle quark distributions. Gluons and light quarks
($u,d$) are considered as massless point-like particles, while for
strange quarks ($s,\overline{s}$) we consider a finite mass
$m_s=150$ MeV.

\section{Mixed hadron-quark-gluon phase}
In this section, we investigate the hadron-quark phase transition
at finite temperature and baryon chemical potential. In a theory with only gluons and no quarks, the transition turns out to be of first
order.  Since the $u$ and $d$ quarks have a small
bare mass, while the strange quark has a somewhat larger mass,
the phase transition is predicted to be a smooth cross-over.
In fact, various results from QCD-inspired models indicate
that, as the baryon chemical potential increases in the phase diagram, a region of non-singular but rapid crossover of thermodynamic observable around a quasi-critical temperature, leads to a critical endpoint (CEP), beyond which the system shows a first-order phase transition from confined to deconfined matter. However, the existence or exclusion of a CEP has not yet been well confirmed by QCD lattice simulations. Since it occurs over a very narrow range of temperatures, the transition, for several practical purposes, can still be considered of first order. Indeed the lattice data with 2 or 3 dynamical flavours are not precise enough to unambiguously disentangle the difference between the two situations.  Thus, by considering the deconfinement transition at finite density as a first-order one, a mixed phase can be formed, which is typically described using the two separate equations of state: one for the hadronic phase, and one for the quark phase.

To describe the mixed phase, we apply the Gibbs formalism to systems with more than one conserved charge \cite{glen2,andrea}, requiring that baryon number, electric charge and strangeness number are preserved.  The main result is that, at variance with the so-called Maxwell construction, the pressure in the mixed phase is not strictly constant and therefore, for instance, the nuclear incompressibility does not vanish.

The structure of the mixed phase is obtained by imposing the Gibbs conditions for chemical potentials ($\mu_B^{(H)} = \mu_B^{(Q)}$, $\mu_C^{(H)} = \mu_C^{(Q)}$, $\mu_S^{(H)} = \mu_S^{(Q)}$) and pressure ($P^H=P^Q$) and by requiring global conservation of the total baryon, electric charge and strangeness densities in the hadronic (H) and quark (Q) phases:
\begin{eqnarray}
&&P^{(H)}(T,\mu_B,\mu_C,\mu_S)=P^{(Q)}(T,\mu_B,\mu_C,\mu_S)\, ,
\nonumber \\
&&\rho_B=(1-\chi)\rho_B^H+\chi \rho_B^Q\,, \nonumber \\
&&\rho_C=(1-\chi)\rho_C^H+\chi \rho_C^Q \, ,\nonumber \\
&&\rho_S=(1-\chi)\rho_S^H+\chi \rho_S^Q \, ,
\end{eqnarray}
where $\chi$ is the fraction of quark-gluon matter in the mixed phase.  In this way we can find the phase coexistence region in the $(T,\mu_B,\mu_C,\mu_S)$ space. At fixed $T$ and $\mu_B$, the charge $\mu_C$ and strangeness $\mu_S$ chemical potentials are obtained by fixing the total electric charge $Z/A$ (for example, $Z/A=0.4$ for lead-lead heavy ion collisions) and the total strangeness neutrality by the conditions
\begin{eqnarray}
\rho_C=\frac{Z}{A}\,\rho_B \, , \ \ \ \rho_S=0 \, .
\end{eqnarray}

In Fig. 1, we report the variation of the pressure as a function of baryon density (in units of nuclear saturation density $\rho_0= 0.153$ fm$^{-3}$) by considering the degrees of freedom of strange matter in the hadronic and quark phase, for different values of $q$. The temperature is fixed at $T = 120$ MeV and the electric charge fraction to $Z/A = 0.4$. In presence of non-extensive effects the pressure results to be considerably increased even for small deviations from the extensive standard statistics.
\begin{figure}[t]
\begin{center}
\resizebox{0.5\textwidth}{!}{%
\includegraphics{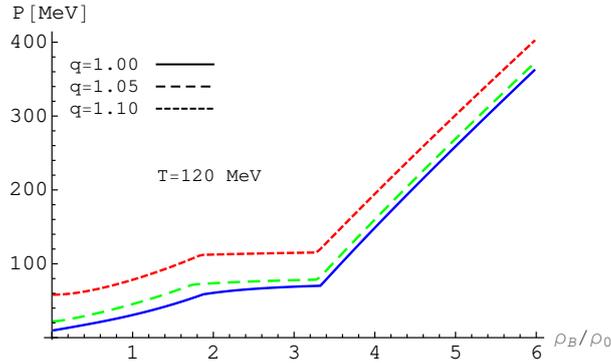}
} \caption{Pressure as a function of the baryon density for different values of non-extensive parameter $q$ (GM3 parameters set).} \label{fig:1}
\end{center}
\end{figure}

\begin{figure}[h]
\begin{center}
\resizebox{0.5\textwidth}{!}{%
\includegraphics{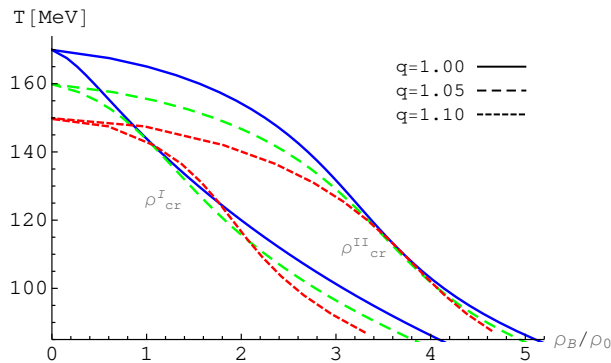}
} \caption{Phase diagram in the $\rho_B -T$ plane for different values of $q$. The curves labeled with $\rho^{I}_{cr}$ and $\rho^{II}_{cr}$ stand for the beginning and the end of the mixed phase,
respectively.} \label{fig:2}
\end{center}
\end{figure}
It is interesting to observe that pressure (as a function of baryon density) is stiffer in the pure hadronic phase, but a strong softening appears in the mixed phase. This feature is remarkable evident by increasing the non-extensive entropic parameter $q$; it implies an abrupt variation in the incompressibility and may be particularly important in identifying the presence of non-extensive effects in high-energy compressed baryonic matter experiments.  Indirect indications of a sensible softening of the EOS at the energies reached at AGS have already been raised elsewhere \cite{iva,bonanno,sahu,stocker,isse}.

In Fig. 2, we show the phase diagram in the $\rho_B -T$ plane for different
values of $q$. The curves labeled with $\rho^{I}_{cr}$ and $\rho^{II}_{cr}$ denote, respectively, the beginning and the end of the mixed phase. In presence of non-extensive statistical effects, we observe a remarkable lowering of the critical maximum temperature at vanishing baryon density $\rho_B$.

In Fig. 3, we show the most relevant particle concentration $Y_i$ as a function of the baryon density in the pure hadron phase, mixed phase and quark phase, at fixed temperature $T=120$ MeV for $q=1$ (left panel) and $q=1.1$ (right panel). As we can see, in presence of non-extensive statistical effects, the particle concentrations clearly differ. In particular, we observe a strong reduction of the neutron and proton fractions with a consistent enhancement of the hyperon and meson fractions, whereas the quark concentrations are not sensibly modified.
%This aspect could be very relevant in the phenomenological interpretation of the relativistic heavy %ion collisions data.
%
\begin{figure}[t]
\begin{center}
\resizebox{1.0\textwidth}{!}{%
\includegraphics{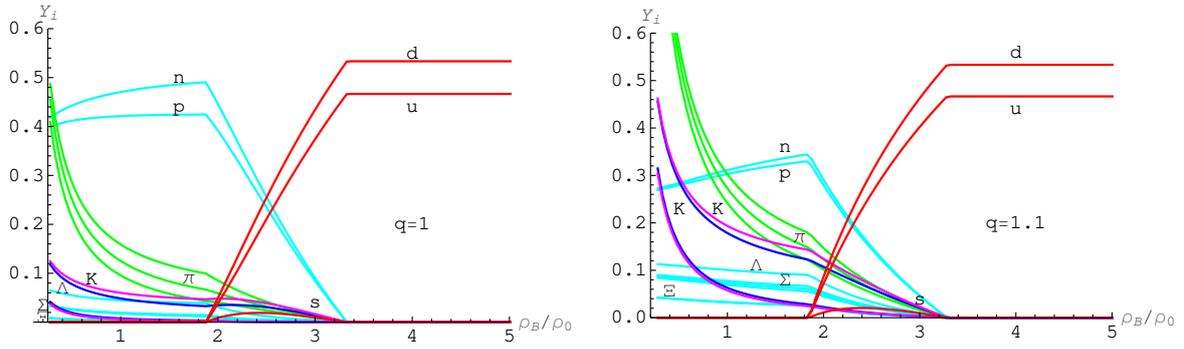}
} \caption{Particle concentrations as a function of the baryon density in the pure hadronic phase, mixed phase and quark-gluon matter for $q=1$ (left panel) and $q=1.1$ (left panel), $T=120$ MeV and $Z/A = 0.4$.} \label{fig:3}
\end{center}
\end{figure}
\begin{figure}[h]
\begin{center}
\resizebox{0.5\textwidth}{!}{%
\includegraphics{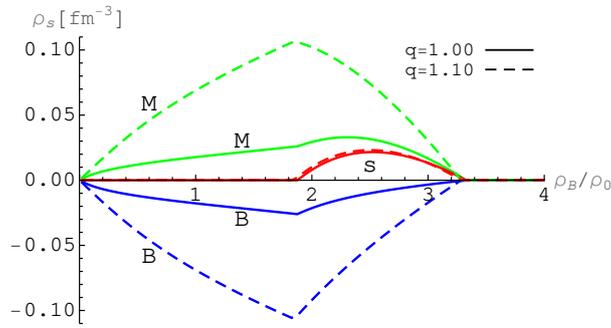}
} \caption{Net strangeness densities of baryons (B), mesons (M) and strange quarks
(s) for different values of the non-extensive parameter $q$ at a fixed temperature
of $T=120$ MeV, in pure hadronic phase, mixed phase and quark-gluon phase.} \label{fig:4}
\end{center}
\end{figure}

Finally, in Fig. 4, we show the net strangeness density $\rho_s$ for the different particle species (baryons, mesons and strange quarks) as a function of baryon
density at fixed temperature $T = 120$ MeV and for different values of $q$. In
presence of non-extensive statistical effects ($q = 1.1$), we observe a strong enhancement
of the strange hadronic particles production, especially in the pure hadronic phase.
At the beginning of the mixed phase ($\rho_B=2\rho_0$), we can observe a maximum in the baryon and mesons net strangeness density, of about $4$ times the values obtained in the standard statistical case ($q=1$), after which it decreases until zero at the beginning of the quark phase. Nonextensive statistical effects have instead a lower impact on the quark sector and the variation of the quark strange density results to be negligible.

\begin{figure}[ht]
\begin{center}
\resizebox{0.5\textwidth}{!}{%
\includegraphics{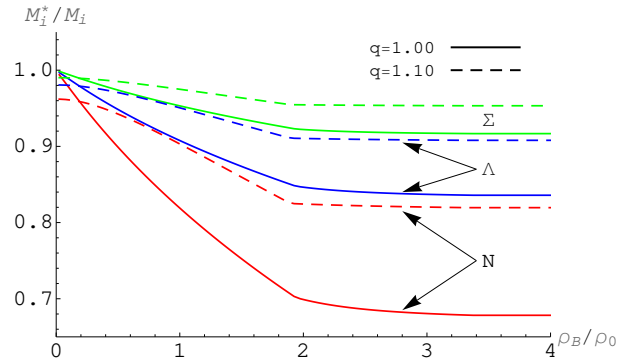}
} \caption{Effective masses ratios ($M^*_i/M_i$) of the most relevant baryon particles (nucleons N, $\Lambda$, $\Sigma$) versus the baryon density $\rho_B$, for different values of the non-extensive parameter $q$ at a fixed temperature $T=120$ MeV.} \label{fig:5}
\end{center}
\end{figure}

To better understand the enhancement of the strange hadronic particle in presence of non-extensive statistical effects observed in the previous two figures, in Fig. 5 we report the effective mass $M^*_i$ (with respect to the corresponding vacuum mass $M_i$) for the most relevant baryons (nucleons N, $\Lambda$, $\Sigma$) as a function of the baryon density at the same values of $q$ and temperature reported in Fig.s 3 and 4.

The behavior of the effective baryon mass as a function of the baryon density $\rho_B$ (or baryon chemical potential $\mu_B$) is related to the behavior of the $\sigma$ field by means of the relation $M^*_i=M_i-g_{\sigma i}\sigma$ (see Section 2). According to the results of Ref. \cite{1lavoro}, we find that at low  $\rho_B$, in presence of non-extensive effects, the value of the $\sigma$ meson field is increased respect to the standard case and decreased at higher values of $\rho_B$. This important feature is due to the fact that for $q>1$ and fixed baryon
density (or $\mu_B$), the (normalized) mean occupation function is enhanced at high values of its argument and depressed at low values. Being the argument of the mean occupation function
$x_i=\beta (E_i^*-\mu_i^*)$, in the integration over momentum (energy), at lower $\mu_{B}$ (corresponding to lower values of the effective particle chemical potential $\mu_i^*$) the enhanced
Tsallis high energy tail weighs much more that at higher $\mu_{B}$ where depressed low energy effects prevail and the mean occupation number results to be bigger for the standard Fermi-Dirac
statistics. Concerning the antiparticle contribution, the argument of $\overline{n}_i$ is $\overline{x}_i=\beta (E_i^*+\mu_i^*)$ and the Tsallis enhancement at high energy tail is favored also at higher $\mu_B$. At the same time, higher temperatures (where antiparticle contribution are more relevant) reduce the value of the argument of $n_i$ and $\overline{n}_i$, favoring the extensive
distribution. These effects are much more evident for the scalar density $\rho_S$ (self-consistently related to the $\sigma$ meson field) where appears $(n_i)^q$ and particle and antiparticle
contributions are summed.
As a consequence, the baryon effective masses become, respect to the standard case, smaller for lower values of $\rho_{B}$ and bigger for higher values.
As can be seen in Fig. 5, this feature appears to be more significant as a percentage for lighter baryons, therefore, at large baryon densities ($\rho_B\approx 1\div 2 \,\rho_0$), the nucleons effective mass is enhanced respect to the standard statistical case ($q=1$) as a sensible greater percentage with respect to the $\Lambda$ and $\Sigma$ effective masses. This matter of fact favors the formation of hyperons respect to nucleons at a fixed baryon density and in presence of non-extensive statistical effects. Moreover, being the strangeness number globally conserved, an increase of hyperon particles implies, in the pure hadron phase, a corresponding increase of strange meson particles in order to satisfy the condition of zero net strangeness.
This effect is also present in the mixed hadron-quark-gluon phase because of the strange quark formation is not sensible affected by non-extensive statistical effects (such a behavior could be due to the simple MIT bag model adopted in this paper; in this sense it could be very interesting future investigations with more sophisticated SU(3) chiral quark models).

In light of the above results, a significative enhancement of the strange to non-strange particle ratios at finite temperature and baryon density could be an evident signature of the relevance of non-extensive statistical effects in the future compressed baryonic matter experiments.

\end{document}